\documentclass[aip,graphicx,amsmath,amssymb,reprint]{revtex4-1}
\usepackage{graphicx}
\usepackage{dcolumn}
\usepackage{bm}
\usepackage[utf8]{inputenc}
\usepackage[T1]{fontenc}
\usepackage{mathptmx}
\usepackage{etoolbox}
\usepackage{subcaption}
\usepackage{tabulary}
\usepackage{color}
\usepackage{hyperref}

\newcommand{\appropto}{\mathrel{\vcenter{
  \offinterlineskip\halign{\hfil$##$\cr
    \propto\cr\noalign{\kern2pt}\sim\cr\noalign{\kern-2pt}}}}} 

\makeatletter
\def\@email#1#2{%
 \endgroup
 \patchcmd{\titleblock@produce}
  {\frontmatter@RRAPformat}
  {\frontmatter@RRAPformat{\produce@RRAP{*#1\href{mailto:#2}{#2}}}\frontmatter@RRAPformat}
  {}{}
}%
\makeatother

\draft 

\begin{document}


\title{Demonstration of a 1820 channel multiplexer for transition-edge sensor bolometers}

\author{J.C. Groh}
 \email{john.groh@lbl.gov}
\affiliation{Quantum Sensors Division, National Institute of Standards and Technology, Boulder, CO 80305}
\affiliation{Physics Division, Lawrence Berkeley National Laboratory, Berkeley, CA 94720}

\author{Z. Ahmed}
\affiliation{Kavli Institute of Particle Astrophysics and Cosmology, Menlo Park, CA 94025}
\affiliation{SLAC National Accelerator Laboratory, Menlo Park, CA 94025}

\author{J. Austermann
}
\affiliation{Quantum Sensors Division, National Institute of Standards and Technology, Boulder, CO 80305}

\author{J. Beall}
\affiliation{Quantum Sensors Division, National Institute of Standards and Technology, Boulder, CO 80305}

\author{D. Daniel}
\affiliation{Department of Physics, University of Colorado Boulder, Boulder, CO 80309}

\author{S.M. Duff}
\affiliation{Quantum Sensors Division, National Institute of Standards and Technology, Boulder, CO 80305}

\author{S.W. Henderson}
\affiliation{Kavli Institute of Particle Astrophysics and Cosmology, Menlo Park, CA 94025}
\affiliation{SLAC National Accelerator Laboratory, Menlo Park, CA 94025}

\author{J. Hubmayr}
\affiliation{Quantum Sensors Division, National Institute of Standards and Technology, Boulder, CO 80305}

\author{R. Lew}
\affiliation{Theiss Research, La Jolla, CA 92037}

\author{M. Link}
\affiliation{Quantum Sensors Division, National Institute of Standards and Technology, Boulder, CO 80305}

\author{T.J. Lucas}
\affiliation{Quantum Sensors Division, National Institute of Standards and Technology, Boulder, CO 80305}

\author{J.A.B. Mates}
\affiliation{Quantum Sensors Division, National Institute of Standards and Technology, Boulder, CO 80305}

\author{M. Silva-Feaver}
\affiliation{Department of Physics, Yale University, New Haven, CT 06520, USA}

\author{R. Singh}
\affiliation{Department of Physics, University of Colorado Boulder, Boulder, CO 80309}
\affiliation{Quantum Sensors Division, National Institute of Standards and Technology, Boulder, CO 80305}

\author{J. Ullom}
\affiliation{Quantum Sensors Division, National Institute of Standards and Technology, Boulder, CO 80305}
\affiliation{Department of Physics, University of Colorado Boulder, Boulder, CO 80309}

\author{L. Vale}
\affiliation{Quantum Sensors Division, National Institute of Standards and Technology, Boulder, CO 80305}

\author{J. Van Lanen}
\affiliation{Quantum Sensors Division, National Institute of Standards and Technology, Boulder, CO 80305}

\author{M. Vissers}
\affiliation{Quantum Sensors Division, National Institute of Standards and Technology, Boulder, CO 80305}

\author{C. Yu}
\affiliation{Department of Physics, University of Chicago, Chicago, IL 60637}
\affiliation{Argonne National Laboratory, Lemont, IL 60439}


\begin{abstract}
This article may be downloaded for personal use only. Any other use requires prior permission of the author and AIP Publishing. This article appeared in \textit{Appl. Phys. Lett.} 127, 152602 (2025), and may be found at \url{https://doi.org/10.1063/5.0290914}.\\

The scalability of most transition-edge sensor arrays is limited by the multiplexing technology which combines their signals over a reduced number of wires and amplifiers.  In this Letter, we present and demonstrate a multiplexer design optimized for transition-edge sensor bolometers with 1820 sensors per readout unit, a factor of two more than the previous state-of-the-art.  The design is optimized for cosmic microwave background imaging applications, and it builds on previous microwave superconducting quantum interference device multiplexers by doubling the available readout bandwidth to the full 4--8 GHz octave.  Evaluating the key performance metrics of yield, sensitivity, and crosstalk through laboratory testing, we find an end-to-end operable detector yield of 78\%, a typical nearest-neighbor crosstalk amplitude of $\sim$0.4\%, and a median white noise level of 83 pA/$\sqrt{\mathrm{Hz}}$ due to the multiplexer, corresponding to an estimated contribution of 4\% to the total system noise for a ground-based cosmic microwave background telescope.  Additionally, we identify a possible path toward reducing resonator loss for future designs with reduced noise. We expect these developments to alleviate the system complexity, cryogenic requirements, and cost of future large arrays of low temperature detectors.

\end{abstract}

\maketitle 


\par
Large arrays of transition-edge sensors (TESs) are increasingly useful for a wide variety of applications including the calorimetric detection of single photons over a variety of wavelengths \cite{LitaReview}, charged particles \cite{ReviewLTDNeutrinoMass}, and low-energy quasiparticle excitations \cite{KimSuperconductingReview} as well as background-limited bolometric measurements of electromagnetic radiation from millimeter to far-infrared wavelengths \cite{HubmayrCMBDetectorReview, FarrahFIRReview}.  A common limitation of such arrays is that the number of sensor elements is constrained by the ability to multiplex their signals over a reasonable number of wires spanning from the cryogenic stage to the room temperature electronics.  The developments presented in this Letter target ground-based mm-wave imaging telescopes for observations of the cosmic microwave background (CMB), though we expect them to be of broad utility.  Current and near-term CMB observatories incorporate arrays of $10^4$--$10^5$ TES bolometers  \cite{CLASSSPIE2014, SobrinSPT3G, HuiBA, ade2019simons, salatino2020design}, and future projects anticipate growth by at least another order of magnitude \cite{abazajian2019cmb, aiola2022snowmass2021}.

\par
Several multiplexing technologies have been developed which are relevant for TES bolometers observing the CMB.   A number of time-division \cite{HendersonACTTDM} and MHz frequency division \cite{BenderSPT3GReadout} systems have been used in CMB telescopes, but these techniques are fundamentally limited in their channel handling capacity by their $\sim$10 MHz total bandwidth: a multiplexing factor of 128 is the largest demonstrated with either method \cite{piat2022qubic}.

To increase the total readout bandwidth, a number of microwave techniques are being advanced which couple the sensor signals to the resonance frequencies of an array of superconducting $\sim$GHz resonators, modulating part of the resonator inductance through either a changing kinetic inductance contribution \cite{KPUP,MKING,KICS} or a changing Josephson inductance in a superconducting quantum interference device (SQUID) \cite{irwin2004microwave}.  The SQUID-based microwave multiplexing technology is now quite mature, and has seen use in a number of telescopes \cite{MUSTANG2, Keckumux, mccarrick2021simons, salatino2020design} with a multiplexing factor as high as 910.  In this Letter, we report on the demonstration of a microwave SQUID multiplexer which increases the multiplexing factor to 1820 without requiring any improvements to the chip microfabrication process, the detector and readout module assembly, the cryogenic cabling and amplification, or the room temperature electronics. This advancement allows future experiments to halve the number of cryogenic coaxial cables and cryogenic amplifiers, thereby reducing system cost and complexity while also relaxing the requirements on cryocoolers.

\par
\begin{figure*}
    \centering
    \includegraphics[width=0.8\textwidth]{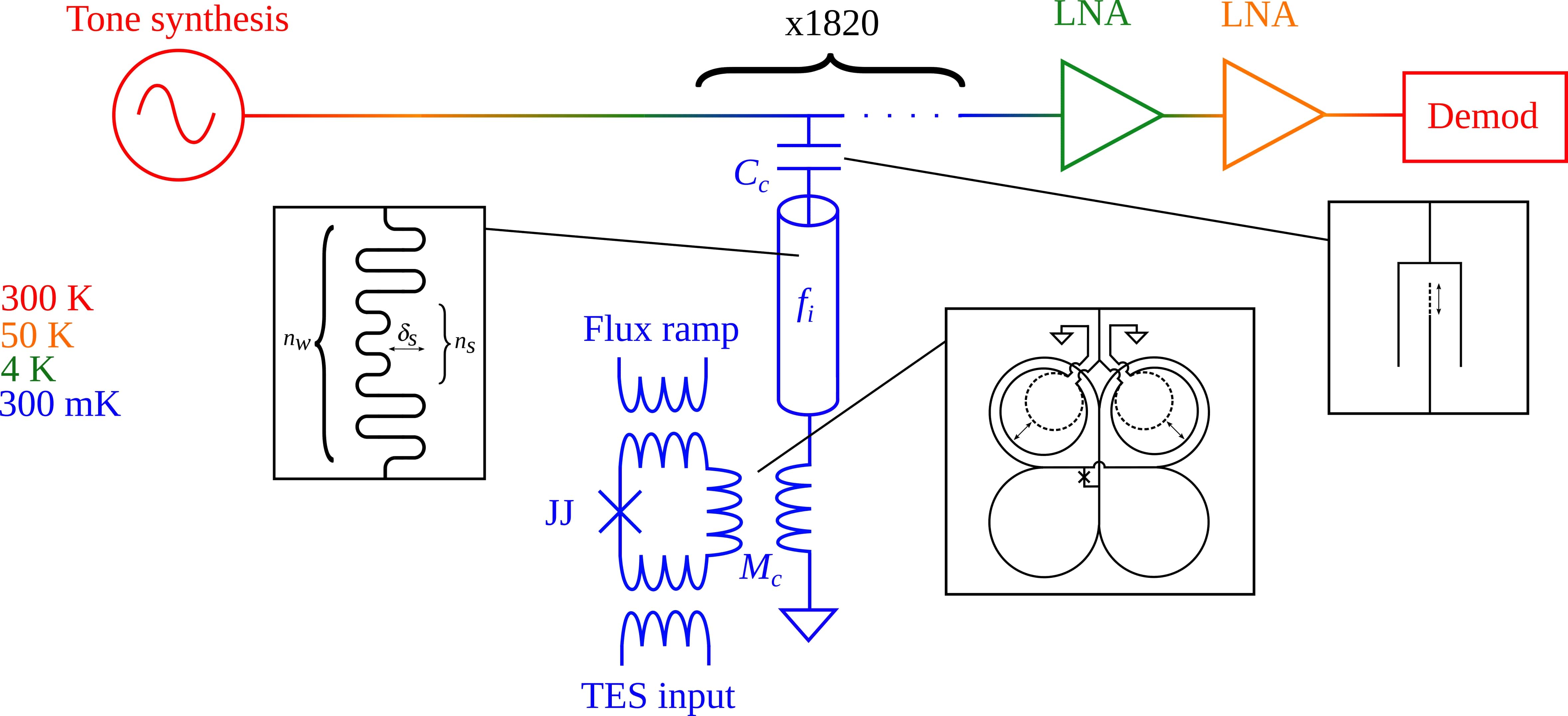}
    \caption{Schematic representation of the microwave SQUID multiplexer which also depicts the general method for lithographically adjusting the resonator frequency and its couplings to the SQUID and feedline.   The strength of the capacitive coupling $C_c$ between the resonator and the feedline is adjusted via the finger length in an interdigital capacitor, and the strength of the inductive coupling $M_c$ between the resonator termination and the SQUID is tuned by adjusting the overlap area of their respective inductive loops.  The resonator spacings are tuned through the number of ``wiggles'' $n_w$, the number of ``sliders'' $n_s$, and the indent of the sliders $\delta_s$.  In the example resonator shown, $n_w = 11$ and $n_s = 2$.  The ranges these circuit parameters take are given in Table \ref{tab:parameters}.}
    \label{fig:schematic}
\end{figure*}

In our microwave SQUID multiplexer, depicted schematically in Figure~\ref{fig:schematic}, each sensor induces a changing magnetic flux in an rf SQUID, which in turn modulates the resonance frequency of a quarter-wave resonator.  An additional common ``flux ramp'' linearizes the system by encoding the detector signals in the phases of the modulated resonator positions.  Many resonators with unique resonance frequencies are capacitively coupled to a common microwave feedline.  A set of probe tones corresponding to the resonator channel frequencies interrogate the resonator positions in transmission.  Further details on the design principles, operation, and implementation of microwave SQUID multiplexers have been reported elsewhere \cite{MatesThesis, KempfMMCumux, dober2021umuxbolometer, mccarrick2021simons}.

\par
The total bandwidth accessible to a microwave SQUID multiplexer is limited by 2$^\mathrm{nd}$-order intermodulation products, which must be avoided by restricting the probe tones to a single octave.  We place our resonators within the 4--8 GHz octave to overlap with commercially available  microwave components, maintain a small chip size, and utilize recently developed custom tone-tracking electronics \cite{yu2023slac}.  Previous microwave SQUID multiplexers have been demonstrated operating only from 4--6 GHz \cite{mccarrick2021simons}; this work therefore doubles the operating bandwidth and the multiplexing factor without increasing the average channel density.

\par
\begin{figure*}
    \centering
    \includegraphics[width=\textwidth]{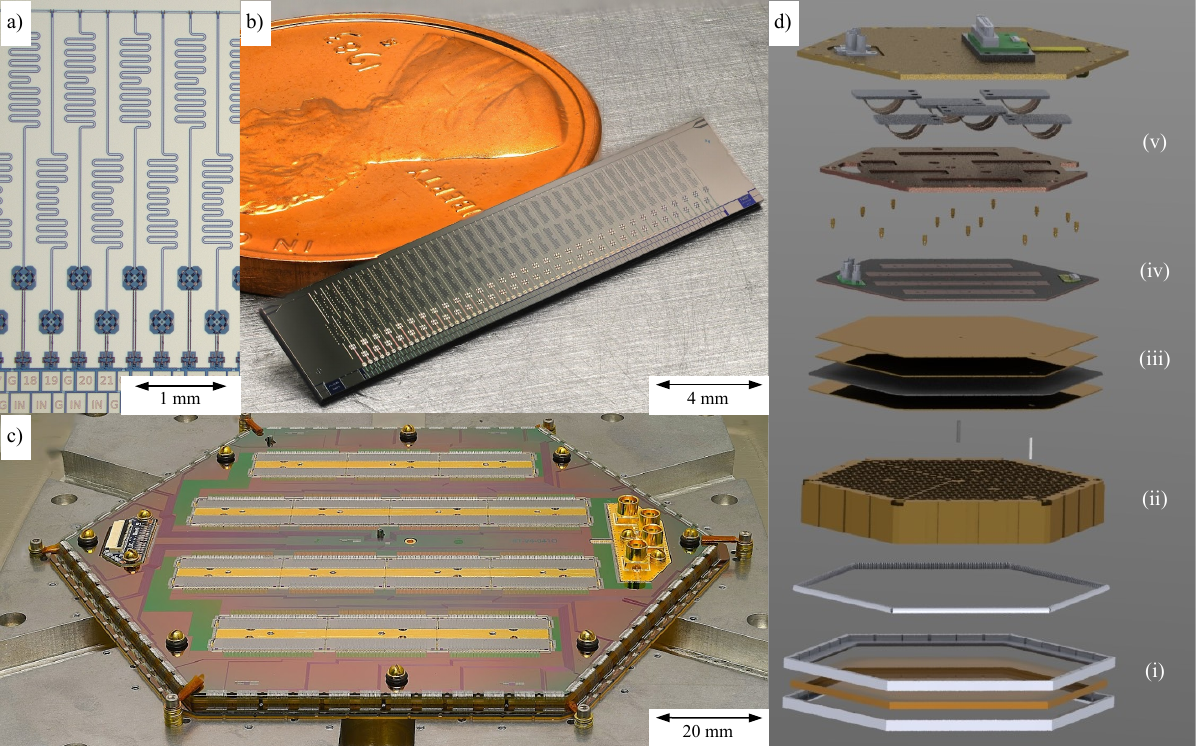} 
    \caption{Physical implementation of the multiplexer.  \textit{a)} Micrograph of a few microwave SQUID channels.  \textit{b)} Photograph of an individual chip.  \textit{c)} Photograph of the 28-chip multiplexer assembly, integrated vertically with the TES wafer via wire bonds along its perimeter.  \textit{d)} Exploded CAD model of the full detector and readout module, showing (i) an optical low-pass filter \cite{MetalMeshFilters}, (ii) a feedhorn array\cite{SiliconPlateletFeedhornArray}, (iii) the stack of optical coupling wafers and the TES wafer, (iv) the multiplexer assembly, and (v) the spring-loaded copper lid which encloses the resonators in an electrically small volume.}
    \label{fig:hardware}
\end{figure*}

We modified only the multiplexer chips containing the SQUIDs and resonators, using existing cryogenic assembly hardware \cite{salatino2020design}.  Enabled by the small readout chip size, this assembly places the readout components directly behind the detector wafer and allows for the efficient tiling of many modules to populate a telescope's focal plane.  Figure \ref{fig:hardware} shows the demonstration hardware setup, which features 28 unique 4 mm $\times$ 20 mm multiplexer chips, each containing 65 channels and 1 "bare" calibration resonator, of an updated design -- described subsequently in this Letter --integrated with an array of TES bolometers and their associated feedhorn coupling optics into a modular package compatible with existing CMB projects currently under construction \cite{ade2019simons,salatino2020design}.  The detector array consists of polarization sensitive TES bolometers designed for operation at a bath temperature of 0.3 Kelvin and with passbands centered at 90 GHz and 150 GHz, similar to those in the AliCPT experiment \cite{salatino2020design}, though 100 mK detectors are also compatible with the multiplexer. The integrated detector and readout module requires two coaxial cables for the common feedline along with 13 twisted wire pairs to carry the flux ramp and TES biases.  

\par
Two principal improvements have been made in the multiplexer chip design relative to previous work\cite{dober2021umuxbolometer}.  First, we identified a parasitic on-chip microwave resonance which became excited by readout tones $\gtrsim$7~GHz, interfering with the operation of higher frequency channels.  This was mitigated straightforwardly by increasing the trace spacing between the input wires leading to the SQUID, reducing their pairwise capacitance and pushing the parasitic resonance safely above the 8~GHz upper end of the readout band.  Second, we have optimized the coupling between each resonator to its SQUID and to the common feedline to better control readout noise and crosstalk across the full 4--8~GHz frequency range.  These couplings are described presently.

\par
\begin{figure*}
    \begin{center}
    \includegraphics[width=\textwidth]{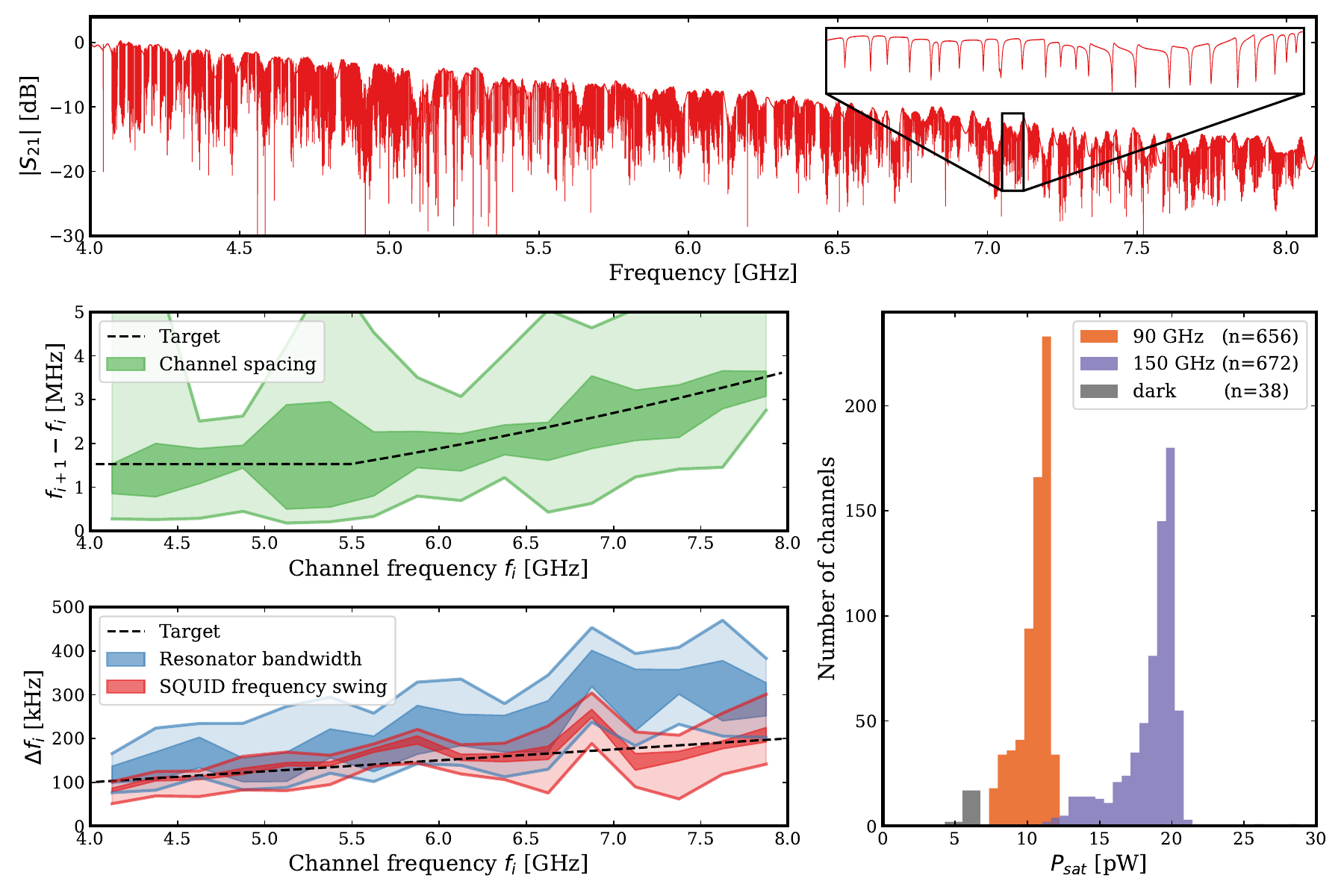} 
    \end{center}
    \caption{Low-level characterization of the fully assembled multiplexer.  \textit{Top:} Measured readout line transmission for the fully integrated readout and detector module, normalized at 4 GHz and measured with a fixed sweep power that corresponds to $\approx$-70 dBm on the mux chip feedline at 6 GHz.  \textit{Center left:} Target and measured pairwise channel spacings, extracted from the above transmission measurement. \textit{Bottom left:} Target and measured channel bandwidths, measured using the SMuRF readout system under nominal operating conditions.  \textit{Bottom right:} Measured saturation powers of 1366 out of a maximum possible 1748 detectors, extracted from measurements of the TES I-V relationship.  In all panels, dark and light shaded bands contain the central 50\% and 90\% of the measured channels in each frequency bin, respectively.}
    \label{fig:operability}
\end{figure*}

Previous work has identified two primary noise mechanisms in the multiplexed microwave SQUID readout currently used for CMB detectors: noise due to the synthesis of the microwave probe tones, and noise due to resonator two-level system (TLS) fluctuations \cite{silva2023microwave}. Assuming tone tracking readout, the equivalent TES current noise $i_n$ due to a voltage noise $v_{n,synth}$ on the feedline induced by the synthesis chain is
\begin{eqnarray}    
    i_{n,synth} & \approx & \frac{\Phi_0Q_cf_r}{2\pi Q^2M_{in}V_{tone}\Delta f_{pp}}v_{n,synth}\label{eq:dac_noise_1}\\
    & \approx & \frac{\Phi_0}{2\pi M_{in}V_{tone}}\left(1 + \frac{1}{Q_i}\frac{f_r}{\Delta f_{BW}}\right)v_{n,synth}
    \label{eq:dac_noise_2}
\end{eqnarray}
where $\Phi_0$ is the magnetic flux quantum, $Q$ is the total quality factor, $Q_c$ is the quality factor of the coupling to the feedline, $f_r$ is the average resonator frequency, $M_{in}$ is mutual inductance of the sensor input to the SQUID, $V_{tone}$ is the amplitude of the voltage oscillation on the feedline due to the probe tone, $\Delta f_{pp}$ is the peak-to-peak swing of the resonator frequ
ency, and $\Delta f_{BW}$ is the resonator bandwidth.  The value and trend with microwave frequency of $v_{n,synth}$ depends on the implementation of the room temperature electronics as well as the flux ramp modulation frequency.  Similarly, the equivalent TES current noise corresponding to a fractional frequency noise $S_{TLS}$ from TLS fluctuations is
\begin{equation}
    i_{n,TLS} = \frac{\sqrt{2}\Phi_0f_r}{\pi M_{in}\Delta f_{pp}}S_{TLS}^{1/2} \propto \frac{f_r}{\Delta f_{BW}}S_{TLS}^{1/2}
    \label{eq:tls_noise}
\end{equation}
The value of $S_{TLS}$ depends on many factors such as tone power, temperature, flux ramp modulation frequency, and the details of the resonator fabrication \cite{zmuidzinas2012}.  Equations (\ref{eq:dac_noise_2}) and (\ref{eq:tls_noise}) show that the main noise contributions scale linearly with frequency and inversely with bandwidth. To counteract the scaling with resonator frequency, we implement a microwave frequency-dependent resonator bandwidth, ranging from 100 kHz at 4 GHz to 200 kHz at 8 GHz (shown on the bottom left in  Figure~\ref{fig:operability}), to maintain low noise across the full readout band.  We note that the signals of interest in ground-based CMB telescopes are $\lesssim$100~Hz, so the modified resonator bandwidth remains independent of the signal bandwidth. The adjustable bandwidth is accomplished by tuning the strength of the capacitive coupling $C_c$ between the resonator and the feedline as shown in Figure~\ref{fig:schematic}. The peak-to-peak modulation of the resonance frequency must also be matched to the resonator bandwidth, so we also adjust the strength of the inductive coupling $M_c$, also shown in Figure~\ref{fig:schematic}.  Neither adjustment requires a change in the fabrication procedure.  The ranges of values which $C_c$ and $M_c$ take across the multiplexer are shown in Table~\ref{tab:parameters}.

\begin{table*}
\centering
\begin{tabular}{|c|c|c|c|}
   \hline
    High-level parameter & Range & Relevant low-level parameter(s) & Range\\
    \hline
   $\Delta f_{BW}$ & 100$-$200 kHz & $C_c$ & 2.2$-$4.2 fF\\ 
    \hline
    $\Delta f_{pp}$ & 100$-$200 kHz & $M_c$ & 1.2$-$1.7 pH\\
    \hline
     &  & $n_w$ & 5$-$21\\
    $f_{i+1} - f_i$ & 1.5$-$3.6 MHz & $n_s$ & 1$-$4\\
     & & $\delta_s$ & 4$-$107 $\mu$m\\
    \hline
\end{tabular}
\caption{Design ranges for key parameters in the multiplexer.  Those not listed here are held at the values specified in previous work \cite{dober2021umuxbolometer}.}
\label{tab:parameters}
\end{table*}

\par
The multiplexer chip design also targets a reduction of inter-channel crosstalk.  The coupling of neighboring resonance frequencies has been found to scale strongly with the average channel frequency $\bar{f}\equiv (f_i+f_{i\pm 1})/2$ in addition to their spacing \cite{groh2024crosstalk, mates2019crosstalk}
\begin{equation}
    \frac{df_i}{df_{i\pm 1}} \propto \frac{\bar{f}^4}{(f_{i\pm 1} - f_i)^2}
\end{equation}
To mitigate the rise at high frequencies, we deviate from the previous design paradigm of constant pairwise channel spacing to a frequency-dependent spacing schedule.  However, fabrication-induced frequency scatter means that scheduling the spacing as $\bar{f}^2$ over the full 4--8 GHz range would result in insufficient margin in the pairwise spacing at the low edge of the band.  Thus, we compromise by fixing the spacing to a constant value of 1.5 MHz at low frequencies, then allow it to rise quadratically for $\bar{f} > 5.5$ GHz (shown in the center left of Figure~\ref{fig:operability}).  The resonator spacings are set by tuning the electrical lengths of their transmission line sections as depicted in Figure~\ref{fig:schematic}.

\par
To evaluate the performance of this design and demonstration assembly, we performed a series of cryogenic tests.  First, we cooled individual multiplexer chips, which enabled us to directly inject signals with dedicated wiring and measure crosstalk.  Next, we integrated the 28-chip multiplexer assembly together to measure its behavior without the complicating influence of detectors.  Finally, we integrated the multiplexer with a TES array and blanked-off feedhorns to measure end-to-end performance with negligible optical power incident on the TESs.  All tests were performed at a temperature of 300 mK, unless otherwise noted, with cryogenic attenuation and amplification components representative of that used in upcoming CMB experiments \cite{RaoSOPlumbing, salatino2020design}.  All measurements involving flux ramp modulated resonances use the SLAC Microresonator RF (SMuRF) tone-tracking electronics \cite{yu2023slac} for tone generation and demodulation with a flux ramp waveform of 5~$\Phi_0$ amplitude and 4~kHz frequency, resulting in a net modulation rate of 20~kHz.  As the signals of interest for CMB measurements are $\lesssim$100~Hz, we downsample the raw 4~kHz data stream by a factor of 20 before writing to disk.

\par
We focus on verifying aspects of the multiplexer which are either fundamentally new or which may have changed with the updated design.  Previous work has demonstrated that the multiplexer component fabrication and module assembly is scalable, high yield, and produces consistent noise performance across the $>10^5$ channels produced for the Simons Observatory \cite{whipps2023high, jones2024qualification, dutcher2024simons, satterthwaite2024SOumux, Healy2022modulerobustness, Healy2022UHFUFM, Li2020moduleassembly}.  The readout noise with the SMuRF electronics and very similar chips has been shown to be white above $\sim$10 mHz with a rising $1/f$-like component at lower frequencies. \cite{henderson2018smurf, mccarrick2021simons}.  The magnetic sensitivity of similar multiplexer chips and assemblies has also previously been quantified \cite{ConnorsMagnetic, Huber2022magnetic, Vavagiakis2021magnetic, Vavagiakis2018magnetic}.

\par
\begin{figure*}
    \begin{center}
    \includegraphics[width=\textwidth]{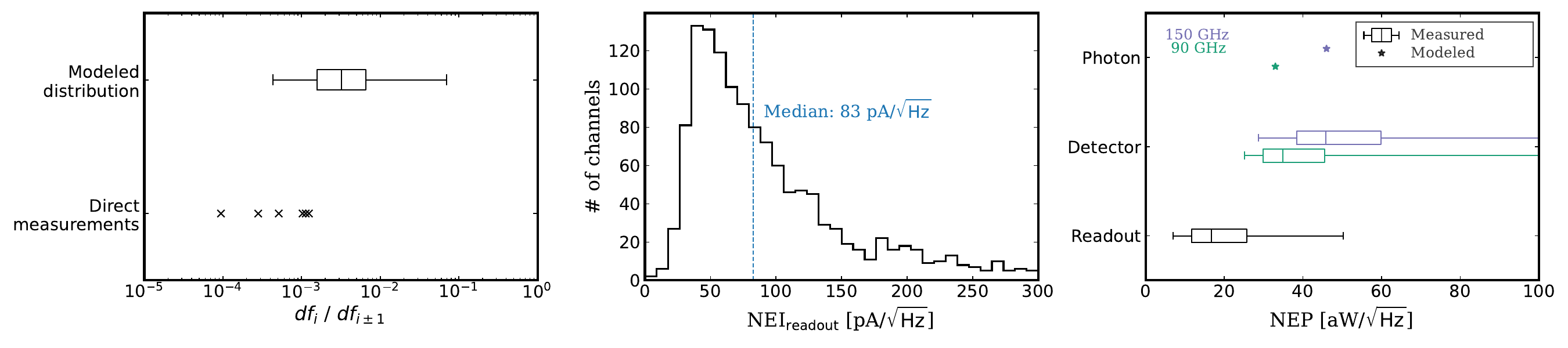} 
    \end{center}
    \caption{Key performance metrics.  \textit{Left:} Measured nearest-frequency-neighbor crosstalk amplitudes for 3 channel pairs from single-chip tests, along with the modeled distribution across all channels in the multiplexer.  The direct measurements are consistent with expectations given the particular spacing of each pair.  We model that 82\% of neighboring channel pairs will experience crosstalk <$10^{-2}$.  \textit{Center:} Measured white noise level of the readout, referenced to an equivalent current fluctuation through the TES.  \textit{Right:} Measured distributions of the readout and detector thermal fluctuation   white noise contributions to the total noise, referenced to an equivalent power fluctuation measured by the TES bolometer.  For further comparison, projected photon noise levels for an example CMB telescope \cite{salatino2020design} are also shown.  In all panels, the boxes and whiskers contain the central 50\% and 90\% of channels.}
    \label{fig:sensitivity}
\end{figure*}

Evaluating first the resonators across the full 4--8 GHz readout band using a vector network analyzer, we find 97\% of the maximum possible 1848 resonances, the transmission of which is shown in the top panel of Figure~\ref{fig:operability}.  We extract the pairwise frequency spacing of these resonators and find concordance with our design as shown in the center left panel of Figure~\ref{fig:operability}. We estimate that 4\% of neighboring resonators are inoperable due to resonance collisions. Inserting these resonator frequencies and pairwise spacings into a crosstalk model \cite{groh2024crosstalk}, the dominant crosstalk mechanism for 98\% of channel pairs is due to the electrical coupling of frequency-adjacent resonators, and the expected median nearest frequency neighbor crosstalk amplitude $df_i/df_{i\pm 1}$ is $3.2\times 10^{-3}$, comparable with that expected in the 910-channel multiplexers in the Simons Observatory as well as typical crosstalk amplitudes in alternative TES multiplexers \cite{groh2024crosstalk, SobrinSPT3G, ade2022bicep3instrument}.  The expected distribution is shown on the left in Figure~\ref{fig:sensitivity}.  As an additional check, we measured the crosstalk amplitude between 3 channel pairs by directly injecting electrical signals during the individual chip tests; these crosstalk amplitudes are also shown on the left in Figure~\ref{fig:sensitivity}.


\par
We next evaluate the multiplexer under nominal operating conditions with the SMuRF system, with tone power levels optimized for minimum noise every 500 MHz.  The SQUID response and resonator bandwidth, shown in the bottom left of Figure~\ref{fig:operability}, largely follow the design, though the resonator bandwidths are $\sim$1.5x larger than expected due to anomalously low $Q_i$ (discussed later).  We are able to characterize, bias, and otherwise operate the TES bolometers without issue, measuring high-quality I-V relationships on 1366 (78\%) out of a maximum possible 1748 detectors, as shown by the histograms of saturation powers for the 3 types of detectors on the wafer in the bottom right of Figure~\ref{fig:operability}.  Modest yield losses in this R\&D assembly were observed due to a variety of factors, including wire bond and trace issues, resonator collisions, and inoperable detectors.  These were not studied in detail, as higher yields have already been demonstrated in very similar hardware with a scaled-up production line. For example, the Simons Observatory has reported an operable yield of 84\% from its production units \cite{dutcher2024simons}.

\par
To estimate the readout contribution to the noise, we heat the bath temperature to $T$ = 600 mK (above the detector $T_c$ of 485 mK) to remove the contribution from thermal fluctuation noise across the bolometer weak link, record 30 seconds of data, compute the power spectral density, and evaluate the median value between 5 Hz and 50 Hz - a region which is both visually white and also safely separated from the pulse tube motor and mains frequencies - of the resulting spectral density.  Because Johnson noise due to the resistive TES is non-negligible at 600 mK, we subtract in quadrature from the measured white noise level a modeled contribution of $i_{n,Johnson} = \sqrt{4k_BT/(R_N+R_{sh})}$, using uniform values of $T$ = 600 mK, $R_n$ = 8 m$\Omega$, and $R_{sh}$ = 400 $\mu\Omega$ for all channels.  The distribution of these estimated readout noise levels with all channels simultaneously running is shown in the center panel of Figure~\ref{fig:sensitivity}, and it has a median value of 83 pA/$\sqrt{\mathrm{Hz}}$.  We find no notable correlation between the readout noise and resonator frequency outside of a few narrow-band features which coincide with features in the SMuRF tone generation noise.

\par
To evaluate the impact of the readout on total sensitivity, we compare to the other two key noise sources: thermal fluctuation noise in the bolometer weak link and photon statistics.  We measure the detector thermal fluctuation noise at a bath temperature of 300 mK with the detectors biased to a median $R/R_n$ = 0.54(0.58) for the 90(150) GHz detectors, removing the relatively small measured contribution from the readout in quadrature.  The distributions of these measured detector and readout noise contributions are shown in the right panel of Figure~\ref{fig:sensitivity}, where we have assumed the high loop gain limit for each detector's power-to-current responsivity: $dI/dP = -1/(I_{TES}(R_{TES}-R_{sh}))$.  We also show estimated photon noise levels for the AliCPT-1 observatory \cite{salatino2020design} as an example relevant application.  As readout, detector, and photon noise are uncorrelated, they add in quadrature and the net impact on sensitivity is small.  Our multiplexer would increase the total NEP by only 3.6\% over a hypothetical experiment with no readout noise.

\par
\begin{figure*}
    \begin{center}
    \includegraphics[width=\textwidth]{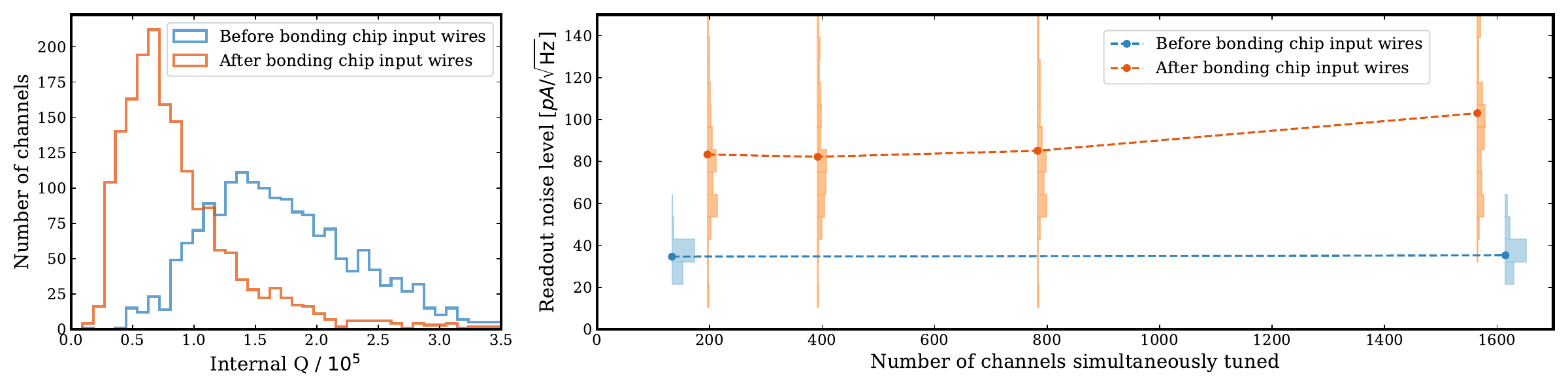} 
    \end{center}
    \caption{\textit{Left:} Measured $Q_i$ of the resonators under nominal power before and after wire bonding the inputs of the multiplexer chips to the detectors and intermediate routing traces.  All other wire bonds (flux ramp, feedline, and perimeter ground bonds) were present for both measurements.  \textit{Right:} Measured readout noise from a set of 100 channels before wire bonding the inputs of the multiplexer chips to the detectors and intermediate routing traces, along with another set of 100 channels measured afterwards.  The shaded regions are histograms of per-channel the readout white noise level, and dots indicate the median white noise level for each configuration.  After connecting the multiplexer chip inputs, the noise level is elevated due to the reduced $Q_i$ and becomes slightly dependent on the number of channels simultaneously operating due to the impact of 3$^{\mathrm{rd}}$-order intermodulation distortion products.}
    \label{fig:degradation}
\end{figure*}

As mentioned earlier, the internal quality factors of the resonators are lower than expected.  In fact, $Q_i$ was observed to be $\sim$2x larger during testing without detectors, degrading only after wire bonding the inputs of the multiplexer chips to the detectors and intermediate wiring traces as shown on the left in Figure~\ref{fig:degradation}.  Subsequent bond removal tests showed that any electrical connection to the signal input coils of each mux channel is sufficient to induce this effect.  Combined with the observation that a few resonators not coupled to SQUIDs also consistently have higher $Q_i$ and do not change with the bonding configuration of adjacent channels, we conclude that the loss mechanism results from unmodeled microwave effects in the coupling circuitry between the TES input coil and the resonator.  The impact of the reduced $Q_i$ is twofold: it reduces the readout sensitivity to TES signals, elevating the overall noise level, and it enhances a sub-dominant contribution to the total noise level from the forest of 3$^{\mathrm{rd}}$-order intermodulation products.  Both effects are visible in the right panel of Figure~\ref{fig:degradation}.   Future designs may be able to achieve lower noise by mitigating this source of resonator loss. Possible avenues include performing a dedicated study to identify a more optimal wiring layout which does not induce as much resonator loss, reducing the sensitivity to $Q_i$ by decreasing $Q_c$ (trading reduced noise for either increased crosstalk or a reduced multiplexing factor), and/or improving the ground distribution via vertical spring-loaded pins.

\par
To conclude, we have developed a multiplexer which can read out 1820 TES bolometers per unit, a full factor of two above the previous state-of-the-art.  The multiplexer is optimized for ground-based CMB telescope applications, and requires no improvements to the chip microfabrication or readout electronics.  Its readout noise is strongly sub-dominant to the photon noise for ground-based CMB telescopes, and the amplitude of inter-channel crosstalk is comparable to other multiplexing technologies.  This development was achieved by increasing the usable readout bandwidth to the full 4--8~GHz octave.  Key technical advances included fixing a design error which caused a parasitic on-chip resonance, and adjusting the coupling strengths $C_c$ and $M_c$ to maintain good noise and crosstalk at 8 GHz.  We also investigated the limiting factor for the readout noise, identifying possible directions for future performance improvements.  We expect the developments described in this Letter to alleviate the system complexity, cryogenic requirements, and cost of future large arrays of low temperature detectors, potentially including major projects under design and construction such as the Advanced Simons Observatory \cite{abitbol2025simons}, AliCPT \cite{salatino2020design}, and SPT-3G+, as well as ambitious proposed projects such as CMB-HD \cite{sehgal2019cmb}.

\section*{supplementary material}
See supplementary material for the derivation of Equations~(\ref{eq:dac_noise_1})--(\ref{eq:tls_noise}).
\begin{acknowledgments}
JCG was supported in part by a fellowship through the National Research Council Research Apprenticeship Program. The authors at NIST acknowledge the support of the NASA APRA program.  The work at SLAC was supported by the Department of Energy under contract DE-AC02-76SF00515.
\end{acknowledgments}

\bibliography{refs_1820x.bib}


\end{document}